\documentclass[fleqn,usenatbib]{mnras}

\usepackage{newtxtext,newtxmath}

\usepackage[T1]{fontenc}

\DeclareRobustCommand{\VAN}[3]{#2}
\let\VANthebibliography\thebibliography
\def\thebibliography{\DeclareRobustCommand{\VAN}[3]{##3}\VANthebibliography}

\usepackage{graphicx}	
\usepackage{amsmath}	

\title[HD 22064]{Fundamental effective temperature measurements for eclipsing binary stars - IV. Selection of new benchmark stars and first results for HD 22064.}

\author[P. F. L. Maxted]{
Pierre F. L. Maxted\thanks{E-mail: p.maxted@keele.ac.uk}  \\
$^{1}$Astrophysics group, Keele University, Staffs, ST5 5BG, UK\\
}

\date{Accepted XXX. Received YYY; in original form ZZZ}

\pubyear{2015}

\begin{document}
\label{firstpage}
\pagerange{\pageref{firstpage}--\pageref{lastpage}}
\maketitle

\begin{abstract}
 I describe the selection and initial characterisation of 20 eclipsing binary stars that are suitable for calibration and testing of stellar models and data analysis algorithms used by the PLATO mission and spectroscopic surveys. 
 The binary stars selected are F-/G-type dwarf stars with M-type dwarf companions that contribute less than 2\,per~cent of the flux at optical  wavelengths. The light curves typically show well-defined total eclipses with very little variability between the eclipses.
 I have used near-infrared spectra obtained by the APOGEE survey to measure the spectroscopic orbit for both stars in HD~22064. Combined with an analysis of the TESS light curve, I derive the following masses and radii: $M_1 = 1.35 \pm 0.03 M_{\odot}$, $M_2 = 0.58 \pm 0.01 M_{\odot}$, $R_1 = 1.554  \pm 0.014  R_{\odot}$, $R_2 = 0.595  \pm 0.008 R_{\odot}$. 
 Using $R_1$ and the parallax from Gaia EDR3, I find that the primary star's angular diameter is $\theta = 0.1035 \pm 0.0009 $\,mas. 
 The apparent bolometric flux of the primary star is ${\mathcal F}_{\oplus,0} = (7.51\pm 0.09)\times10^{-9}$\,erg\,cm$^{-2}$\,s$^{-1}$. 
 Hence, this F2V star has an effective temperature $T_{\rm eff,1} = 6763{\rm\,K} \pm 39{\rm \,K}$. 
 HD~22064 is an ideal benchmark star that can be used for ``end-to-end'' tests of the stellar parameters measured by large-scale spectroscopic surveys, or stellar parameters derived from asteroseismology with PLATO. 
 The techniques described here for HD~22064 can be applied to the other eclipsing binaries in the sample in order to create an all-sky network of such benchmark stars.
\end{abstract}



\section{Introduction}
Apart from the Sun and a few nearby stars, detached eclipsing binaries (DEBS) are our best source of fundamental data for normal stars. Accurate, model-independent mass and radius measurements with a precision of $\pm 0.5$\,per~cent or better are now feasible thanks to the availability of space-based photometry and high-quality echelle spectroscopy \citep{2020MNRAS.498..332M}. The advent of precise parallax measurements from GAIA now also makes it possible to measure the effective temperatures of both stars in an eclipsing binary to $\pm50$\,K or better directly from their angular diameters and bolometric fluxes \citep{2020MNRAS.497.2899M}. Well-studied DEBS typically have flux ratios in the optical $L_2/L_1 \approx 1$, i.e. they are SB2 systems at optical wavelengths.  This makes it straightforward to measure the radial velocity for both stars. These DEBS are excellent benchmark stars for testing stellar models but, in general, cannot be used for ``end-to-end'' tests of stellar spectroscopy pipelines designed to measure the atmospheric parameters of single stars. It is sometimes possible to obtain a spectrum for one star if the system shows total eclipses, but scheduling these observations can be difficult, particularly for long-period DEBS.

Echelle spectrographs operating in the near-infrared now make it possible to make direct mass, radius and effective temperature (T$_{\rm eff}$) measurements for DEBS with optical flux ratios $\ell \ll 1$\,per~cent. This has been demonstrated for EBLM~J0113+31 \citep{2022MNRAS.513.6042M}. These systems are ideal benchmark stars for the PLATO stellar spectroscopy and asteroseismology pipelines because they look like single stars at optical wavelengths. However, the telescope time required to measure model-independent mass, radius and T$_{\rm eff}$ for systems with extreme flux ratios like EBLM J0113+31 is expensive and difficult obtain, e.g. a 4-sigma detection of the M-dwarf in EBLM~J0113+31 required 22 spectra obtained with a spectrograph on a 3.6-m telescope. However, it is not necessary to go to such extremes to create suitable benchmark stars. DEBS with flux ratios $\ell\approx 1$\,per~cent are much easier to characterise in detail, while the flux from the M-dwarf will have a little effect on the stellar parameters derived from spectroscopy at optical wavelengths \citep{2018MNRAS.473.5043E}. The secondary star in these DEBS can be characterised in some detail, so it is feasible to remove the signal of the M-dwarf from the combined spectrum of the binary, leaving a ``clean'' spectrum of the primary star suitable for detailed spectroscopic analysis.   There is certainly no impact from the M-dwarf on the asteroseismic signal that will be measured for the primary star in these binaries by PLATO mission \citep{2014ExA....38..249R}. This makes these DEBS ideal candidates for testing the stellar models used by the PLATO mission, and for validation of the stellar parameters that will be provided in the PLATO mission data products \citep{2022A&A...658A.147G}.

I have therefore selected 20 DEBS with V=9-12 and $\ell \approx 1$\,per~cent at optical wavelengths having light curves showing narrow, total eclipses and little or no variation between the eclipses due to star spot activity or tidal distortion of the primary star.  This sample of DEBS will substantially improve the coverage and quality of benchmark FGK dwarf stars compared to existing benchmark stars \citep{2018RNAAS...2..152J}. These DEBS are 5-10 magnitudes fainter than the existing benchmark stars, which are typically naked-eye stars like $\alpha$~Cen. This puts them in the same magnitude range as stars that can are observed by large-scale spectroscopic surveys using standard observing modes, i.e., these benchmark DEBS can be used for end-to-end tests for all these instruments and data analysis pipelines, enabling us to put the results from these surveys and the PLATO pipeline onto a homogeneous effective temperature scale. 

\section{Target selection}

 Targets have been selected from the Kepler eclipsing binary catalogue\footnote{\url{http://keplerebs.villanova.edu/}} \citep{2016AJ....151...68K}, the TESS eclipsing binary stars catalogue\footnote{\url{http://tessEBs.villanova.edu}} \citep{2019ApJS..244...11K}, \citet{2021ApJ...912..123J} and \citet{2018A&A...616A..38M}. In addition, I consulted various lists of interesting eclipsing binaries I have generated over several years as a result of student projects, inspection of light curves from the WASP survey \citep{pollacco2006}, citizen scientist projects,\footnote{\url{http://www.planethunters.org}} etc. Short period systems (P$<4$ days) were ignored to avoid the complications due to tidal distortions of the stars and rapid rotation. Systems showing little or no variation between the eclipses were preferred, although a few systems showing variations  in brightness $\approx 1$\,per~cent due to star spots were selected so that the sample includes a few stars showing moderate magnetic activity. All the systems selected show a total secondary eclipse with a depth   $\approx 1$\,per~cent and narrow eclipses. Stars with early-type primary stars (A-type stars or hotter), stars with bright companions within a few arcseconds, binaries with a clear detection of third light from an unresolved tertiary star in the light curve,  and stars fainter than $G=13$ were not considered.
The stars selected are listed in Table~\ref{tab:basic_info}.

\section{Methods}

\subsection{Light curve analysis}  
 The typical procedures I used to measure the properties of the selected binary stars from their light curves are described in this subsection. Variations in these procedures for individual systems are described in Section~\ref{sec:notes}. Note that I use the normal convention here for the analysis of eclipsing binary stars of using the term ``eclipse'' to refer to both the primary (deeper) eclipse caused by the transit of the primary star by the secondary star, and the secondary eclipse caused by the occultation of the cooler, smaller secondary star.

 I used {\sc lightkurve}\footnote{\url{https://docs.lightkurve.org/}}  \citep{2018ascl.soft12013L} to search the Mikulski Archive for Space Telescopes\footnote{\url{https://archive.stsci.edu/}} (MAST) for light curves of each binary system observed by either the TESS \citep{2015JATIS...1a4003R}, Kepler \citep{2010Sci...327..977B} or K2 \citep{2014PASP..126..398H} missions. For most of the stars analysed here, I used light curves from the TESS mission observed with a cadence of 120\,s and processed to produce {\sc pdc\_sapflux} values by the TESS Science Processing Operations Center (SPOC). The data were downloaded from MAST using {\sc lightkurve} and bad data rejected using the default bit mask. The depths and widths of the eclipses, and phase of secondary eclipses in each light curve measured from these data are given in Table~\ref{tab:basic_info}. This information plus an an initial estimate of the orbital period and time of primary eclipse were used to identify sections of the light curve containing complete eclipses plus some data either side. These sections of light curve were divided by a straight line fit by least-squares to the data either side of the eclipse, and then exported in a format suitable for analysis using {\sc jktebop}\footnote{\url{http://www.astro.keele.ac.uk/jkt/codes/jktebop.html}} \citep{2010MNRAS.408.1689S}. 

 The NDE light curve model \citep{1972ApJ...174..617N} on which {\sc jktebop} is based computes eclipses using the approximation that the stars are spherical. None of the stars in this sample have an oblateness larger than 0.0015, and the typical oblateness of the primary stars is 0.00003, so this approximation is a very good one for these well-detached binaries. Since the stars and very nearly spherical and we have not used data between the eclipses, the ellipsoidal effect was ignored in the analysis of the light curves. The reflection effect was also ignored. Limb darkening was modelled using the power-2 law recently implemented in {\sc jktebop} \citep{2023arXiv230102531S}. The values of T$_{\rm eff}$, $\log g$ and [Fe/H] for the primary star in Table~\ref{tab:dr3} were used to estimate the values of the parameters $h_1$ and $h_2$ using interpolation within the table from \citet{2018A&A...616A..39M}. The effect of the assumed value of $h_1$ for the primary star can be seen in the curvature of the light curve at the bottom of the primary eclipse so this parameter was allowed to vary in the analysis the light curve. The effect of $h_2$ is far more subtle so this parameter was fixed at the value obtained from  \citet{2018A&A...616A..39M}. The assumed limb darkening of the secondary star has a negligible effect on the light curve so we used the fixed values $h_1=0.6$ and $h_2=0.4$ for the secondary stars in all systems for both TESS and Kepler photometry.

 The parameters of the binary star model are: the sum of the stellar radii in units of
the semi-major axis (fractional radii), $r_1+r_2=(R_1+R_2)/a$; the ratio of the stellar radii, $k=R_2/R_1$; the ratio of the surface brightness at the centre of each stellar disc, $J_0$; the orbital inclination, $i$; the time of mid-primary eclipse, $T_0$; the orbital period, $P$;  $e\sin(\omega)$ and $e\cos(\omega)$, where $e$ is the orbital eccentricity and $\omega$ is the longitude of periastron for the primary star. Least-squares fits obtained by varying all these parameters for each binary star are shown in the supplementary online information that accompanies this article. 

 The time of mid-eclipse for each primary eclipse observed by TESS (including 600-s and 1800-s cadence data) was measured using a similar method but using only the data covering each primary eclipse, and with the values of $P$ and $J_0$ fixed at the values determined from the fit to the whole light curve. For stars observed by the WASP survey \citep{pollacco2006} I measured additional times of mid-eclipse by fitting all the data from each observing season using the same method as for the TESS photometry but with the values of  $r_1+r_2$, $k$ and $i$ fixed at the values measured from the analysis of TESS light curve. The linear ephemerides for the time of primary eclipse obtained from a least-squares fit to all these times of mid-eclipse are given in Table~\ref{tab:ephem}. All the times of mid-primary eclipse used to derive the ephemerides in this table on the BJD$_{\rm TDB}$ time scale are available in the supplementary online information that accompanies this article. I also used a least-squares fit of a quadratic ephemeris to check that there is no significant variation in the measured orbital period for any of the selected binary stars.
 
The best-fit parameters for each light curve are given in Table~\ref{tab:lcfits}. For these results, the orbital period was fixed at the value taken from Table~\ref{tab:ephem} and, where possible, separate least-squares fits were performed on independent sections of the light curve containing at least one primary and one secondary eclipse. The values in Table~\ref{tab:lcfits} are then the mean and standard error of the mean of the best-fit parameters from these independent data sets. In cases where fewer than 5 independent data sets could be constructed from the TESS light curves, the standard error estimates in Table~\ref{tab:lcfits} were estimated using a Monte Carlo method with 1000 trials fitting synthetic data generated from the best-fit light curve model plus Gaussian noise with the same standard deviation as the root-mean-square ({\it rms}) of the residuals from the best fit to the real data (Task 8 in {\sc jktebop}). 

 For EPIC~212822491, EPIC~212801667, EPIC~213843283 and EPIC~206288770, I analysed the light curve derived from images obtained by the K2 mission and corrected for instrumental effects using the {\sc everest} algorithm \citep{2016AJ....152..100L}.\footnote{\url{https://luger.dev/everest/}} The depths of eclipses in the {\sc everest} light curves available from MAST can be shallower than their true depths if the automatic procedure to identify ``outliers'' does not flag these events, and so they are treated as noise by the detrending algorithm. To avoid this problem, I computed new light curves for these stars using functionality available in this  software to mask eclipses ``by hand'' prior to applying the de-trending algorithm. The integration time of 1800\,s used for these light curves was accounted for in the analysis with {\sc jktebop} using numerical integration of the model over 5 points per observation.

\subsection{Mass, radius and effective temperature estimates}
 For binary systems where the semi-amplitude of the primary star's spectroscopic orbit ($K_1$) is available, I have used the empirical relation $M_{\star}({\rm T}_{\rm eff}, {\rm [Fe/H]}, \rho_{\star})$ from \citet{2010A&A...516A..33E} to make an improved estimate of the primary star's mass ($M_1$). The values of T$_{\rm eff}$ and [Fe/H] are taken from Table~\ref{tab:dr3} assuming standard errors of 200\,K and 0.2\,dex, respectively. The mean stellar density of the primary star, $\rho_{\star}$, is determined directly from the value of $R_1/a$ via Kepler's third law using the following equation:
 \begin{equation*}
\mbox{$\rho_{\star}$} =
 \frac{3\mbox{M$_{\star}$}}{4\pi\mbox{R$_{\star}$}^3} = \frac{3\pi}{GP^2(1+q)}
 \left(\frac{a}{\mbox{R$_{\star}$}}\right)^3.
\end{equation*}
To estimate the mass ratio, $q=M_2/M_1$, I used the mass function with the mass estimates from Table~\ref{tab:dr3}. The masses derived from this relation are assumed to have a standard error of 0.023 in $\log_{10}(M_{\star})$, in addition to any uncertainty inherited from the standard errors on the input values. The radius of the primary star then follows directly from the value of $\rho_{\star}$ measured from the light curve. The mass of the companion star can then be estimated from the mass function and its radius from the radius ratio $k=R_2/R_1$ measured from the light curve. 

The average surface brightness ratio, $J = \ell/k^2$, where $\ell$ is the flux ratio, provides a strong constraint on the ratio of the effective temperatures, ${\rm T}_{\rm eff,2}/{\rm T}_{\rm eff,1}$ given some way to estimate the surface brightness of a star as a function of effective temperature, $S({\rm T}_{\rm eff})$. Note that $J_0\ne J$ because limb darkening is stronger for the M-dwarf than for the primary star. I used the spectral energy distributions from \citet{2007MNRAS.382..498C} integrated over the spectral response functions for the TESS and Kepler instruments to determine polynomial relations for $S({\rm T}_{\rm eff})$ and its inverse assuming solar composition for both stars. With these relations, the measured values of $J$ and estimates of ${\rm T}_{\rm eff,1}$ from Table~\ref{tab:dr3}, I was then able to estimate the effective temperature of the companion star, ${\rm T}_{\rm eff,2}$.

\begin{figure*}
	\includegraphics[width=0.8\textwidth]{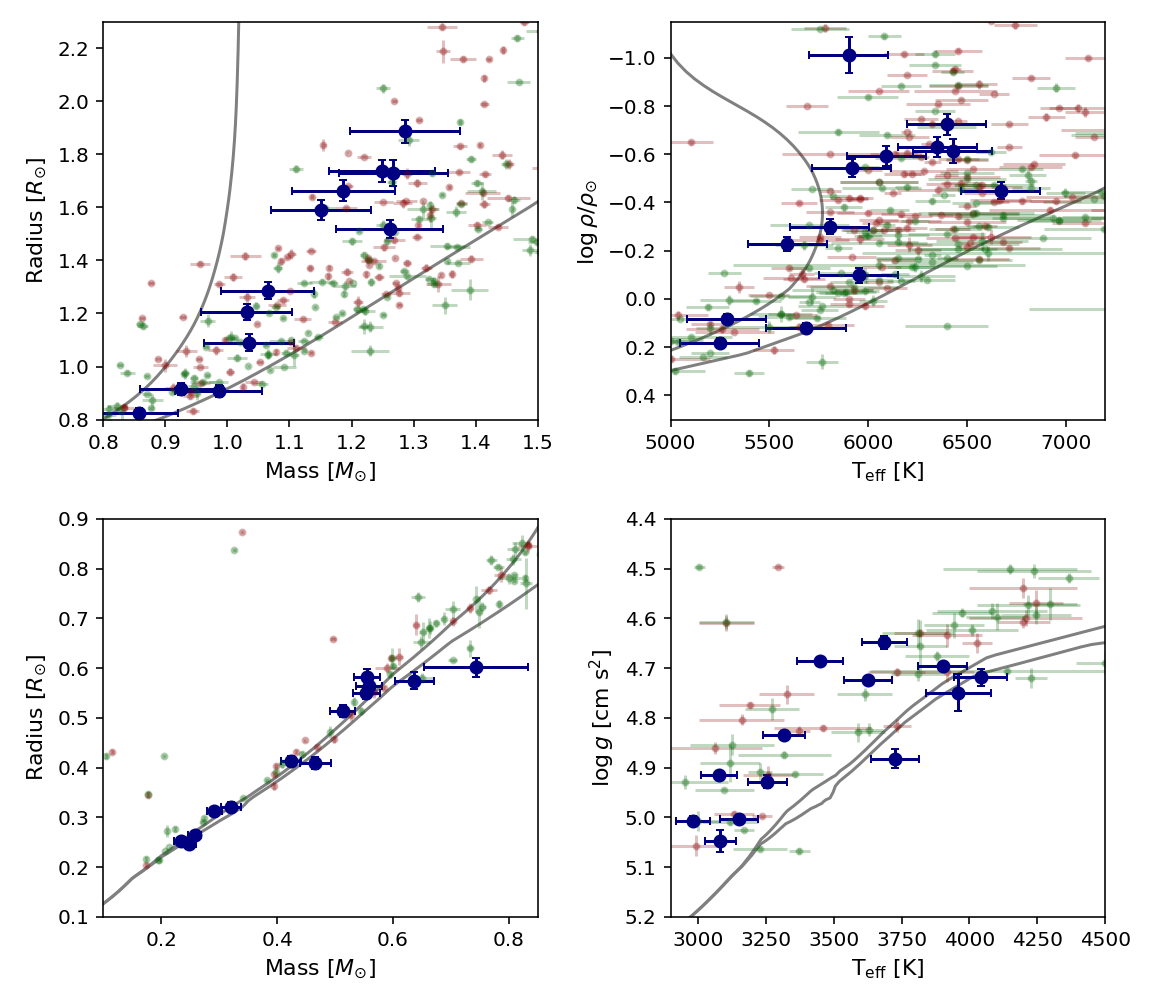}
    \caption{Properties of the primary stars (upper panels) and secondary stars (lower panels) inferred from the light curve analysis and $K_1$ values for the selected binary star systems (blue points with error bars). Red and green points show the properties of primary and secondary stars, respectively, for detached eclipsing binaries taken from DEBcat \citep{2015ASPC..496..164S}. Grey lines are isochrones for solar composition at ages of 1 and 10 Gyr from the MIST grid of stellar models \citep{2016ApJS..222....8D, 2016ApJ...823..102C}. }
    \label{fig:massradius}
\end{figure*}

These mass, radius and T$_{\rm eff,2}$ estimates are given in Table~\ref{tab:massradius} and plotted in Fig.~\ref{fig:massradius}. Also given in Table~\ref{tab:massradius} is the surface gravity of the companion, $\log g_2$, which is independent of the assumed mass of the primary star \citep{2004MNRAS.355..986S}. 

\begin{figure}
	\includegraphics[width=\columnwidth]{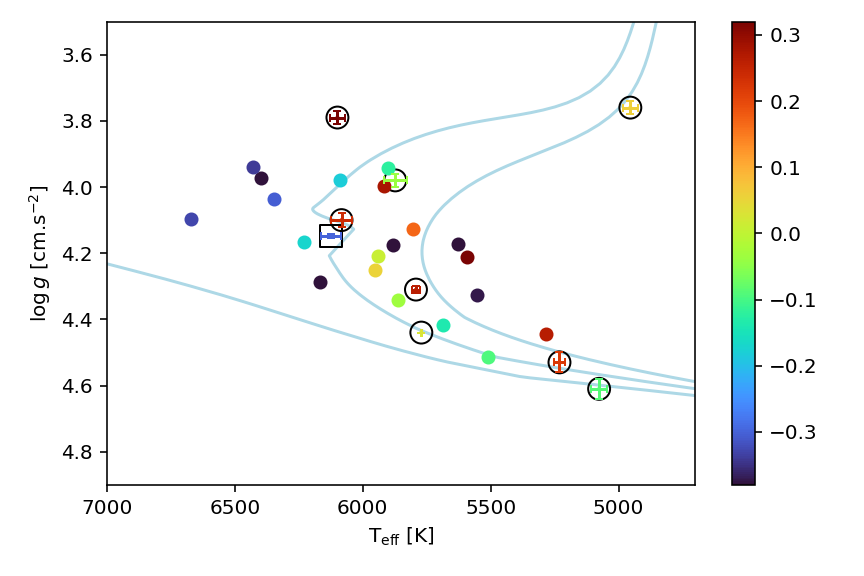}
    \caption{Primary stars in the $T_{\rm eff}$\,--\,$\log g$ plane. Points are colour-coded by [M/H] as indicated in the scale bar. Stars from the Gaia FGK benchmark stars v2.1  catalogue with effective temperature errors $<80$\,K are shown circled and with error bars. Parameters for the benchmark star in the eclipsing binary EBLM~J0113+31 (boxed) are from \citet{2022MNRAS.513.6042M}.  Light blue lines are isochrones for solar composition at ages of 1, 5 and 10 Gyr from the MIST grid of stellar models \citep{2016ApJS..222....8D, 2016ApJ...823..102C}.}
    \label{fig:teff_logg}
\end{figure}

\begin{table*}
	\centering
	\caption{Proposed benchmark eclipsing binary stars. 
 Spectral types (Sp.) are taken from Simbad or estimated based on T$_{\rm eff}$ from Table 2. $D_{1,2}$ and $W_{1,2}$ are the depths and widths (in phase units) of the eclipses. The phase of the secondary eclipse relative to primary eclipse is given in the column $\phi_2$. The semi-amplitude of the primary star's spectroscopic orbit, $K_1$,  is taken from Gaia DR3 \citep{2022yCat.1357....0G} unless otherwise noted.}
\label{tab:basic_info}
\begin{tabular}{lccrlrrrrrrr} 
\hline
Name & $\alpha$ (J2000.0) & $\delta$ (J2000.0) & $G$ [mag] & Sp.& 
\multicolumn{1}{l}{$P$ [d]}& \multicolumn{1}{l}{$D_1$}& 
\multicolumn{1}{l}{ $W_1$}& \multicolumn{1}{l}{$\phi_2$}& 
\multicolumn{1}{l}{$D_2$}& \multicolumn{1}{l}{$W_2$}& 
\multicolumn{1}{l}{$K_1$ [km/s]}\\
		\hline
HD 4875           & 00:50:39.98 & $-$18:30:21.2 &  8.82 & G3V  &  13.64   & 0.12 & 0.023 & 0.464 & 0.011 & 0.016  & $ 29.48 \pm  0.18 $\\
HD 22064          & 03:33:27.57 &   +00:07:10.8 &  8.74 & F2V   &   9.13   & 0.10 & 0.043 & 0.445 & 0.017 & 0.016  &                    \\
CD$-$27 2812        & 06:12:59.66 & $-$27:52:49.4 &  9.63 & F9   &   7.84   & 0.10 & 0.034 & 0.536 & 0.012 & 0.031  & $ 38.71 \pm  0.21 $\\
CD$-$31 3271        & 06:24:48.93 & $-$31:51:52.2 &  9.81 & G0   &   5.61   & 0.19 & 0.030 & 0.500 & 0.015 & 0.030  &                    \\
TYC 8547-22-1     & 06:36:58.94 & $-$58:27:36.6 &  9.96 & G0   &   4.24   & 0.14 & 0.047 & 0.500 & 0.015 & 0.047  & $ 51.31 \pm  0.40 $\\
TYC 8549-603-1   & 06:53:35.69 & $-$58:43:55.3 & 12.31 & G5   &   8.73   & 0.07 & 0.042 & 0.492 & 0.012 & 0.034  & $ 47.89 \pm  4.13 $\\
BD+15 1661        & 07:47:39.71 &   +14:47:44.0 & 10.26 & G0   &  28.14   & 0.08 & 0.012 & 0.300 & 0.005 & 0.011  &                    \\
TYC 3421-1132-1   & 08:17:47.26 &   +51:00:38.1 & 10.44 & K0   &  20.54   & 0.09 & 0.010 & 0.427 & 0.006 & 0.012  &                    \\
TYC 8176-503-1    & 09:44:23.57 & $-$48:50:14.3 & 10.30 & G3   &  11.39   & 0.07 & 0.020 & 0.500 & 0.005 & 0.020  &                    \\
EPIC 212822491    & 13:44:19.44 & $-$04:49:39.0 & 11.04 & G1   &  14.32   & 0.17 & 0.021 & 0.500 & 0.019 & 0.021  &                    \\
EPIC 212801667    & 13:51:58.39 & $-$05:29:27.9 & 11.91 & G7   &  23.27   & 0.09 & 0.011 & 0.593 & 0.004 & 0.016  &                    \\
TYC 7284-224-1$^a$& 14:03:40.19 & $-$32:33:27.2 & 11.96 & G3   &  11.91   & 0.09 & 0.018 & 0.531 & 0.008 & 0.013  & $20.9367 \pm 0.0065$ \\
HD 137267         & 15:28:53.45 & $-$66:29:32.0 &  9.83 &G2IV/V&  13.08   & 0.04 & 0.015 & 0.661 & 0.003 & 0.012  & $ 19.99 \pm  0.47 $\\
TYC 3890-1121-1   & 17:05:25.78 &   +55:43:28.1 & 11.87 & G5   &  23.51   & 0.05 & 0.015 & 0.334 & 0.003 & 0.013  &                    \\
TYC 3538-689-1    & 18:26:24.10 &   +51:16:03.1 & 11.38 & F6   &  18.24   & 0.11 & 0.021 & 0.255 & 0.017 & 0.015  & $ 37.63 \pm  1.06 $\\
EPIC 213843283    & 18:48:21.26 & $-$28:06:37.4 & 11.98 & F7   &  11.22   & 0.18 & 0.024 & 0.501 & 0.015 & 0.023  &                    \\
KOI-7303          & 19:27:44.72 &   +47:18:34.9 & 11.51 & F9   &  13.68   & 0.19 & 0.012 & 0.455 & 0.011 & 0.015  & $ 25.81 \pm  0.67 $\\
KOI-7141$^{b}$    & 19:51:51.87 &   +45:32:44.8 & 11.97 & F6V  &  50.44   & 0.09 & 0.010 & 0.920 & 0.008 & 0.005  & $ 34.13 \pm  0.24 $ \\
TYC 9102-351-1    & 21:25:00.94 & $-$60:21:31.4 & 11.72 & F5   &  31.71   & 0.07 & 0.013 & 0.510 & 0.007 & 0.012  & $ 21.75 \pm  0.73 $\\
EPIC 206288770    & 22:17:03.85 & $-$08:24:47.0 & 12.39 & F8   &  24.76   & 0.19 & 0.011 & 0.679 & 0.014 & 0.015  &                    \\
\hline
\noalign{\smallskip}
\multicolumn{12}{@{}l}{$^a$ $K_1$ from \citet{2019A+A...624A..68M}.} \\
\multicolumn{12}{@{}l}{$^b$ $K_1$ from this study -- see Section \ref{sec:koi-7141}}. \\
\end{tabular}
\end{table*}

\begin{table}
\centering
\caption{Primary star parameters from Gaia DR3 GSP-Phot Aeneas best library using BP/RP spectra. $A_G$ is the estimated extinction in the Gaia $G$ band. The metallicity is calibrated using the routines described in \citep{2022arXiv220606138A}.}
\label{tab:dr3}
 \begin{tabular}{lrrrrr} 
 \hline
 Name & 
 \multicolumn{1}{l}{$T_{\rm eff}$} & 
 \multicolumn{1}{l}{$\log g$}  & 
 \multicolumn{1}{l}{[M/H]} & 
 \multicolumn{1}{l}{$A_G$} &
 \multicolumn{1}{l}{Mass}  \\
&  
\multicolumn{1}{c}{[K]} & 
 \multicolumn{1}{c}{[cm\,s$^{-2}$]} &  & [mag]&  
  \multicolumn{1}{c}{[$M_{\odot}$]}  \\
 \hline
HD 4875          & 5804 & 4.13 &$ +0.17 $&$ 0.001 $&  1.02 \\
HD 22064         & 6670 & 4.10 &$ -0.33 $&$ 0.021 $&  1.38 \\
CD$-$27 2812       & 6091 & 3.98 &$ -0.18 $&$ 0.002 $&  1.29 \\
CD$-$31 3271       & 5861 & 4.34 &$ -0.03 $&$ 0.018 $&  1.01 \\
TYC 8547-22-1    & 5916 & 4.00 &$ +0.28 $&$ 0.001 $&  1.13 \\
TYC 8549-603-1   & 5903 & 3.94 &$ -0.12 $&$ 0.217 $&  1.30 \\
BD+15 1661       & 5941 & 4.21 &$ +0.01 $&$ 0.033 $&  1.07 \\
TYC 3421-1132-1  & 5285 & 4.44 &$ +0.26 $&$ 0.050 $&  0.88 \\
TYC 8176-503-1   & 5592 & 4.21 &$ +0.43 $&$ 0.067 $&  0.92 \\
EPIC 212822491   & 5881 & 4.17 &$ -0.45 $&$ 0.002 $&  1.07 \\
EPIC 212801667   & 5555 & 4.33 &$ -0.37 $&$ 0.002 $&  0.91 \\
TYC 7284-224-1   & 5685 & 4.42 &$ -0.14 $&$ 0.082 $&  0.99 \\
HD 137267        & 5953 & 4.25 &$ +0.05 $&$ 0.052 $&  1.09 \\
TYC 3890-1121-1  & 5628 & 4.17 &$ -0.57 $&$ 0.001 $&  0.99 \\
TYC 3538-689-1   & 6349 & 4.04 &$ -0.30 $&$ 0.093 $&  1.38 \\
EPIC 213843283   & 6229 & 4.17 &$ -0.17 $&$ 0.405 $&  1.06 \\
KOI-7303$^{\dagger}$& 5510 & 4.51 &$ -0.09 $&$ 0.196 $&  0.91 \\
KOI-7141$^{\clubsuit}$& 6397 & 3.97 &$ -0.42 $&$ 0.360 $&  1.47 \\
TYC 9102-351-1$^{\star,\dagger}$ & 6429 & 3.94 &$ -0.34 $&$ 0.041 $&  1.35 \\
EPIC 206288770   & 6165 & 4.29 &$ -0.42 $&$ 0.206 $&  1.08 \\
\hline
\multicolumn{6}{@{}l}{$^{\star}$ Parameters from GSP-Spec \citep{2022arXiv220605541R}.} \\
\multicolumn{6}{@{}l}{$^{\dagger}$ Mass estimate from the TIC.}\\
\multicolumn{6}{@{}l}{$^{\clubsuit}$ No calibration possible for [M/H].}
\end{tabular}
\end{table}

\begin{table}
\centering
\caption{Linear ephemerides for the time of primary eclipse on the BJD$_{\rm TDB}$ time scale.}
\label{tab:ephem}
 \begin{tabular}{lr} 
 \hline
Star & \multicolumn{1}{l}{Ephemeris} \\
 \hline
HD 4875         & $ 2458738.61766(6) +  13.635561(1) \cdot E $ \\ 
HD 22064        & $ 2458578.1295(2) + 9.134881(3) \cdot E$ \\ 
CD$-$27 2812    & $ 2458767.3708(1) + 7.835742(2) \cdot E$ \\ 
CD$-$31 3271    & $ 2458488.46869(2) + 5.6204306(8) \cdot E$ \\ 
TYC 8547-22-1   & $ 2458867.75062(2) + 4.2418215(1) \cdot E$ \\ 
TYC 8549-603-1  & $ 2458630.0641(5) + 8.72983(1) \cdot E$ \\ 
BD+15 1661      & $ 2459036.7160(2) + 28.142049(6) \cdot E$ \\ 
TYC 3421-1132-1 & $ 2459123.71243(9) + 20.544620(3) \cdot E$ \\ 
TYC 8176-503-1  & $ 2458804.51727(5) + 11.394232(2)  \cdot E$ \\ 
EPIC 212822491  & $ 2457942.63157(1) + 14.3206436(4) \cdot E$ \\ 
EPIC 212801667  & $ 2457837.29167(4) + 23.274645(2) \cdot E$ \\ 
HD 137267       & $ 2458794.0346(1) + 13.080733(7) \cdot E$ \\
TYC 3890-1121-1 & $ 2458965.1854(3) + 23.514340(5) \cdot E$ \\
TYC 3538-689-1  & $ 2459197.49620(5) + 18.2416720(6) \cdot E$ \\
EPIC 213843283  & $ 2457302.30170(9) + 11.22161(2) \cdot E$ \\
KOI-7303        & $ 2455378.72159(1) + 13.6836900(2) \cdot E$ \\
KOI-7141        & $ 2456798.7639(5) + 50.44032(2) \cdot E$ \\
TYC 9102-351-1  & $ 2457463.0202(6) + 31.71044(1) \cdot E$ \\
EPIC 206288770  & $ 2457017.79623(5) + 24.756465(3) \cdot E$ \\
\hline
\end{tabular}
\end{table}

\begin{table*}
\centering
\caption{Best-fit parameters from the fits to TESS (T) or Kepler (K) light curves of selected benchmark eclipsing binaries. The value in the column $\ell$ is the flux ratio in the band noted in the second column. Figures in parentheses are the standard error on the final digit of the preceeding value. Where multiple subsets of the light curve have been used to estimate the standard error on the parameters, the number of subsets used is given the column $N_{\rm lc}$.}
\label{tab:lcfits}
\begin{tabular}{lcrrrrrrrrr} 
\hline
Name & Band &  
\multicolumn{1}{c}{$J_0$}& 
\multicolumn{1}{c}{$r_1+r_2$}& 
\multicolumn{1}{c}{$k = r_2/r_1$}& 
\multicolumn{1}{c}{$h_1$}& 
\multicolumn{1}{c}{$i$ $[^{\circ}]$}& 
\multicolumn{1}{c}{$e\cos\omega$}& 
\multicolumn{1}{c}{$e\sin\omega$} &
\multicolumn{1}{c}{$\ell$} & $N_{\rm lc}$  \\
\hline

HD 4875          & T & 0.1100(5)& 0.0619(1)& 0.3212(3)& 0.794(4)& 89.026(7)& -0.05588(3)& -0.173(1)& 0.00967(4)&  \\
HD 22064         & T & 0.123(3)& 0.0940(3)& 0.383(6)& 0.858(7)& 87.82(2)& -0.0732(1)& -0.5320(7)& 0.01471(8)& 5\\
CD-27 2812       & T & 0.1469(8)& 0.1102(4)& 0.3087(3)& 0.822(6)& 87.23(3)& 0.05768(6)& 0.059(3)& 0.01168(5)&  \\
CD-31 3271       & T & 0.100(1)& 0.0986(5)& 0.414(7)& 0.81(2)& 89.4(2)& = 0.0 & = 0.0 & 0.0145(5)& 6\\
TYC 8547-22-1    & T & 0.1130(3)& 0.1631(1)& 0.354(1)& 0.796(2)& 86.20(2)& = 0.0 & = 0.0 & 0.0134(1)& 102\\
TYC 8549-603-1   & T & 0.24(1)& 0.131(2)& 0.2526(5)& 0.854(6)& 88.1(1)& -0.0110(2)& -0.144(8)& 0.0126(7)& 5\\
BD+15 1661       & T & 0.092(1)& 0.0396(2)& 0.2503(3)& 0.798(3)& 89.38(2)& -0.31886(6)& -0.018(4)& 0.00490(6)&  \\
TYC 3421-1132-1  & T & 0.0809(9)& 0.0351(2)& 0.2750(5)& 0.763(4)& 89.85(5)& -0.11492(6)& 0.092(3)& 0.00534(6)&  \\
TYC 8176-503-1   & T & 0.084(2)& 0.0647(2)& 0.2593(5)& 0.829(7)& 88.95(3)& = 0.0 & = 0.0 & 0.00471(8)&  \\
EPIC 212822491   & K & 0.1335(5)& 0.06710(3)& 0.4011(4)& 0.757(2)& 88.462(3)& = 0.0 & = 0.0 & 0.01885(6)& 9\\
EPIC 212801667   & K & 0.0452(3)& 0.0386(1)& 0.2672(2)& 0.748(2)& 89.60(2)& 0.1424(1)& 0.234(3)& 0.00285(2)&  \\
TYC 7284-224-1   & T & 0.089(4)& 0.0497(3)& 0.292(1)& 0.78(1)& 89.62(8)& 0.0468(2)& -0.0979(4)& 0.0065(3)&  \\
HD 137267        & T & 0.068(3)& 0.0527(9)& 0.226(4)& = 0.8 & 88.12(6)& 0.2517(8)& -0.19(2)& 0.00296(8)&  \\
TYC 3890-1121-1  & T & 0.104(3)& 0.0468(2)& 0.2029(4)& 0.802(5)& 89.25(3)& -0.26413(7)& -0.015(4)& 0.0036(1)& 17\\
TYC 3538-689-1   & T & 0.207(3)& 0.06450(9)& 0.331(5)& 0.84(1)& 88.43(3)& -0.3860(5)& -0.218(5)& 0.0188(5)& 6\\
EPIC 213843283   & K & 0.0917(6)& 0.073(1)& 0.3941(7)& 0.776(6)& 89.40(3)& 0.01(1)& -0.003(2)& 0.01301(7)& 6\\
KOI-7303         & K & 0.0775(2)& 0.04514(9)& 0.390(2)& 0.729(6)& 89.74(4)& -0.06978(1)& 0.0999(9)& 0.0106(1)& 19\\
KOI-7141         & K & 0.1015(3)& 0.03481(2)& 0.29187(8)& 0.803(2)& 88.434(2)& 0.6676(1)& -0.4073(3)& 0.00771(3)& 25\\
TYC 9102-351-1   & T & 0.126(4)& 0.0423(6)& 0.236(1)& 0.80(2)& 89.11(5)& 0.0165(1)& -0.03(1)& 0.0065(2)&  \\
EPIC 206288770   & K & 0.0852(2)& 0.04119(7)& 0.4089(3)& 0.763(2)& 89.292(6)& 0.27976(9)& 0.207(2)& 0.01355(4)&  \\

\hline
\end{tabular}
\end{table*}

\begin{table*}
\centering
\caption{Mass and radius estimates for stars in eclipsing binaries with $K_1$ measurements from Gaia DR3.}
\label{tab:massradius}
 \begin{tabular}{lrrrrrrr} 
 \hline
 \multicolumn{1}{l}{Star} &
 \multicolumn{1}{l}{$M_1 [M_{\odot}]$} &
 \multicolumn{1}{l}{$R_1 [R_{\odot}]$} &
 \multicolumn{1}{l}{$M_2 [M_{\odot}]$} &
 \multicolumn{1}{l}{$R_2 [M_{\odot}]$} &
 \multicolumn{1}{l}{T$_{\rm eff,2}$ [K]} &
 \multicolumn{1}{l}{$\log(\rho_1/\rho_{\odot})$} &
 \multicolumn{1}{l}{$\log g_2$ [cm\,s$^{-2}$]} \\
 \hline
HD 4875      &$1.06\pm0.08 $ &$1.29\pm0.03 $&$0.42\pm0.02 $&$0.41\pm0.01 $&$3316 \pm 77 $ &$ -0.299 \pm 0.004 $  &$ 4.835 \pm 0.003 $ \\ 
HD 22064     &$1.26\pm0.09 $ &$1.52\pm0.03 $&$0.55\pm0.02 $&$0.58\pm0.02 $&$3687 \pm 84 $ &$ -0.449 \pm 0.004 $  &$ 4.649 \pm 0.013 $ \\ 
CD-27 2812   &$1.19\pm0.08 $ &$1.66\pm0.04 $&$0.51\pm0.02 $&$0.51\pm0.01 $&$3627 \pm 87 $ &$ -0.593 \pm 0.006 $  &$ 4.725 \pm 0.004 $ \\ 
TYC 8547-22-1 &$1.15\pm0.08 $ &$1.59\pm0.04 $&$0.56\pm0.02 $&$0.56\pm0.01 $&$3450 \pm 83 $ &$ -0.542 \pm 0.005 $  &$ 4.685 \pm 0.005 $ \\ 
TYC 8549-603-1 &$1.32\pm0.09 $ &$2.38\pm0.08 $&$0.74\pm0.09 $&$0.60\pm0.02 $&$3959 \pm 121 $ &$ -1.011 \pm 0.023 $  &$ 4.750 \pm 0.037 $ \\ 
TYC 3421-1132-1 &$0.93\pm0.07 $ &$0.92\pm0.02 $&$0.23\pm0.01 $&$0.25\pm0.01 $&$2983 \pm 63 $ &$ 0.083 \pm 0.006 $  &$ 5.008 \pm 0.010 $ \\ 
TYC 8176-503-1 &$1.03\pm0.07 $ &$1.21\pm0.03 $&$0.29\pm0.01 $&$0.31\pm0.01 $&$3077 \pm 66 $ &$ -0.227 \pm 0.005 $  &$ 4.916 \pm 0.005 $ \\ 
TYC 7284-224-1 &$0.99\pm0.07 $ &$0.91\pm0.02 $&$0.26\pm0.01 $&$0.26\pm0.01 $&$3152 \pm 72 $ &$ 0.120 \pm 0.008 $  &$ 5.003 \pm 0.006 $ \\ 
HD 137267    &$1.03\pm0.07 $ &$1.09\pm0.03 $&$0.25\pm0.01 $&$0.25\pm0.01 $&$3083 \pm 58 $ &$ -0.098 \pm 0.021 $  &$ 5.048 \pm 0.023 $ \\ 
TYC 3538-689-1 &$1.25\pm0.09 $ &$1.74\pm0.04 $&$0.64\pm0.03 $&$0.57\pm0.02 $&$4041 \pm 97 $ &$ -0.630 \pm 0.008 $  &$ 4.719 \pm 0.018 $ \\ 
KOI-7303     &$0.86\pm0.06 $ &$0.82\pm0.02 $&$0.32\pm0.02 $&$0.32\pm0.01 $&$3256 \pm 70 $ &$ 0.183 \pm 0.007 $  &$ 4.928 \pm 0.013 $ \\ 
KOI-7141     &$1.29\pm0.09 $ &$1.89\pm0.04 $&$0.55\pm0.02 $&$0.55\pm0.01 $&$3902 \pm 90 $ &$ -0.725 \pm 0.004 $  &$ 4.695 \pm 0.003 $ \\ 
TYC 9102-351-1 &$1.27\pm0.09 $ &$1.73\pm0.05 $&$0.47\pm0.03 $&$0.41\pm0.01 $&$3724 \pm 88 $ &$ -0.613 \pm 0.019 $  &$ 4.882 \pm 0.019 $ \\

 \hline
\noalign{\smallskip}
\multicolumn{8}{@{}l}{$^{\dagger}$ $K_1$ from fit to radial velocities measured from APOGEE spectra.} \\
\end{tabular}
\end{table*}

\section{Notes on individual systems}
\label{sec:notes}
\subsection{HD 4875}
 Analysis of the Hipparcos astrometry for this binary system and comparison of the proper motion derived from Hipparcos astrometry and Gaia astrometry shows that the proper motion is not constant \citep{2007A&A...464..377F, 2021ApJS..254...42B}. 
 \citet{2019A&A...623A..72K} interpret this proper motion anomaly as orbital motion due to a companion with a mass $\approx 40 M_{\rm Jup}$. 
 A companion of this mass is very likely to be a brown dwarf or, perhaps, a very low mass main-sequence star.
 In either case, the contribution to the observed flux distribution from this third object in the system will be completely negligible, so this does not prevent the selection of this triple system as a benchmark star. The astrometry and radial velocity measurements of this target will be affected by the orbital reflex motion of the binary due to this very low mass companion, so these effects should be accounted for in a detailed analysis of this binary system to derive the mass, radius and effective temperature of the primary and secondary stars. 
 The astrometric solution provided in Gaia DR3 does not account for the orbital motion induced by the tertiary object and so the values of the renormalised unit weight error, {\tt ruwe}=1.815, and the significance of excess noise in the astrometry, {\tt astrometric\_excess\_noise\_sig}=54.12 are larger than expected for a good fit to the data. Gaia data release DR4\footnote{\url{https://www.cosmos.esa.int/web/gaia/release}} will contain all the measurements used to derive astrometric solutions and source classifications, so a complete analysis accounting for the tertiary object using all available data can be done at that time to derive a more accurate parallax for this star.

 \subsection{HD 22064}
 \label{sec:hd22064_lcfit}
  The secondary eclipse in the light curve of this eccentric binary is a total eclipse but the primary is partial. This makes the results from the light curve analysis more sensitive to systematic errors in the depth of the primary eclipses than for other binaries in this sample that have better-defined primary eclipses. An initial analysis of the TESS {\sc pdcsap\_flux} photometry suggested that there is significant third-light in this system. However, this is almost certainly a spurious result caused by systematic errors in the primary eclipse depth caused by the ``pre-search data conditioning'' (PDC de-trending) because the value of $e\cos(\omega)$ obtained from this fit is inconsistent with the value measured from the spectroscopic orbit described in Section~\ref{sec:hd22064}. For the results quoted here, I have used the {\sc sap\_flux} values provided by the TESS SPOC. A small correction was applied to the fluxes in this light curve for the contamination of the photometric aperture by nearby stars using the values of {\sc crowdsap = 0.993} provided in the metadata with the light curve. Fitting this light curve with third light ($\ell_3$) as a free parameter shows that this value is consistent with the assumption $\ell_3=0$. There is very good agreement between the values of  $e\cos(\omega)$ from our adopted light curve solution derived with  $\ell_3=0$ and the fit to the radial velocities described in Section~\ref{sec:hd22064}. 

\subsection{CD-31 3271}
There is a clear signal in the light curve between the eclipses due to star spots on the primary star with a peak-to-peak amplitude up 2~per~cent and a period of about 5.0\,days. I included third light ($\ell_3$) as a free parameter in the light curve analysis as an approximate way to account for unocculted star spots and faculae. The best-fit values of $\ell_3$ vary from $-0.14$ to $0.05$.
 
\subsection{TYC 8547-22-1}
This star has been observed in 25 TESS sectors at 120~s cadence and the orbital period is 4.24 days, so there are more than 250 primary eclipses in the TESS light curve. There is a clear signal in the light curve between the eclipses due to star spots on the primary star with an amplitude of about 0.5~per~cent and a period of about 4.16\,days. As for CD$-$31 3271, $\ell_3$ was included as a free parameter in the light curve analysis. The mean and standard error of these values are $\ell_3=-0.05 \pm  0.08$. No resolved companions were detected in speckle observations of this star by \citet{2021AJ....162..192Z} down to a contrast ratio of 3.05 magnitudes at $0.15\arcsec$ in the I band.



\subsection{TYC 3421-1132-1}
The secondary eclipses of this eccentric binary system were identified as candidate exoplanet transits by the TESS mission, so this star also has a designation as a TESS object of interest -- TOI-1711. Follow-up observations have established that this star is an SB1 binary system, so the star is now correctly flagged as a "false positive" on the  Exoplanet Follow-up Observing Program (ExoFOP) website.\footnote{\url{https://exofop.ipac.caltech.edu/tess/}} As a result of these follow-up observations, high quality spectra at optical wavelengths are available for this star, and high-contrast imaging has ruled out the presence of resolved faint companions or background stars down to a contrast ratio of 4.3 magnitudes in the $z$ band at 1 arcsec separation. This star was also identified as an SB1 binary by \citet{2022AJ....163..297C}. One of the two primary eclipses in the TESS light curve occurs near a gap in the data. This has not been handled well by the PDC de-trending so I have used the {\sc sap\_flux} data for this analysis.

\subsection{TYC 8176-503-1}
 No reliable times of mid-primary eclipse are available from the WASP survey so we supplemented the times of mid-eclipse from TESS with a single measurement based on a fit to the combined V-band data available from the All Sky Automated Survey (ASAS)  survey \citep{1997AcA....47..467P}\footnote{\url{http://www.astrouw.edu.pl/asas/}}  with the parameters $r_1+r_2, k, i$ and $J_0$ fixed to best-fit values from of a preliminary fit to the TESS light curve. The primary eclipse is clearly asymmetric, perhaps due to the companion crossing a star spot on the primary during the transit. There light curve between the eclipses does show variability on a timescale of about 8\,days with a peak-to-peak amplitude of about 1\,per~cent. I did not include third light in the analysis of this light curve because the value obtained has a large uncertainty and is consistent with $\ell_3=0$.
 
\subsection{EPIC 212822491}
 The sixth data release (DR6) of the Radial Velocity Experiment (RAVE) \citep{2020AJ....160...82S} includes estimates of T$_{\rm eff}$, $\log g$ and [Fe/H] obtained using a few different techniques to analyse the single spectrum observed by this survey. These values agree reasonably well with the values in Table~\ref{tab:dr3}.

\subsection{TYC 7284-224-1}
 Monitoring of the radial velocity by the BEBOP project \citep{2019A+A...624A..68M} over a baseline of approximately 3 years shows that the systematic velocity of this binary (EBLM~J1403$-$32) varied by no more than 12\,m\,s$^{-1}$ per year during this time. This makes it very unlikely that there is an unresolved companion star to this eclipsing binary.
 We have used the values of $e$ and $\omega$ from this study to set Gaussian priors on the values of $e\cos(\omega)$ and $e\sin(\omega)$ for the analysis of the TESS light curve.
 \citet{2019A&A...625A.150V} have analysed this star using data from the Swiss Euler 1.2-m telescope. The masses and radii they obtained agree quite  well with the values in Table~\ref{tab:massradius} despite the poor coverage of the transit in their light curve. 
 The second data release of the GALAH survey \citep{2018MNRAS.478.4513B} includes the following estimates for this star:  T$_{\rm eff} = 5658\pm 50$\,K, $\log g=4.40\pm  0.1$, ${\rm [Fe/H]} = -0.19 \pm 0.06$, $[\alpha/{\rm Fe}]= 0.10\pm 0.02$.

\subsection{HD 137267}
This is a similar case to HD~22064, in that the primary eclipse is partial despite the secondary eclipse being total. It is not possible to make a reliable measurement of the limb-darkening parameter $h_1$ for the primary star in this case, so we fixed this parameter at the value obtained from \citet{2018A&A...616A..39M}.  



\subsection{KOI-7303}
 This star has been previously identified as an eclipsing binary using data from the Kepler mission \citep{2016AJ....151...68K, 2019ApJS..244...43Z}. The results in Table~\ref{tab:lcfits}  are from the analysis of the Kepler data using the 60-s cadence {\sc pdcsap\_flux} light curves. The light curve between the eclipses shows variations due to star spots with a period of about 21~days and a peak-to-peak amplitude of up to 2\,per~cent. I included third light ($\ell_3$) as a free parameter in the light curve analysis for KOI-7303 as an approximate way to account for unocculted star spots and faculae. The best-fit values of $\ell_3$ vary from $-0.19$ to $0.04$. 

\subsection{KOI-7141}
\label{sec:koi-7141}
This star has been previously identified as an eclipsing binary using data from the Kepler mission \citep{2016AJ....151...68K, 2019ApJS..244...43Z}. The Kepler {\sc pdcsap\_flux} light curves are clearly affected by problems with the de-trending around the secondary eclipses so I have used the {\sc sap\_flux} light curves for this analysis. A small correction was applied to the fluxes in this light curve for the contamination of the photometric aperture by nearby stars using the value of {\sc crowdsap = 0.997} provided in the metadata with the light curve. 

\begin{figure}
	\includegraphics[width=\columnwidth]{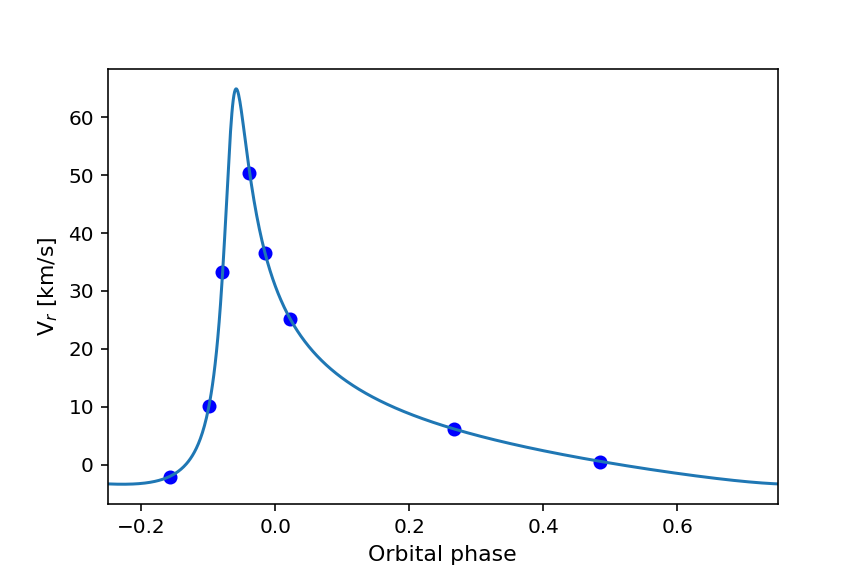}
    \caption{Spectroscopic orbit fit the radial velocities of KOI-7141 measured from spectra obtained with the FIES spectrograph.}
    \label{fig:KOI-7141_rvfit}
\end{figure}

There are 8 spectra of KOI-7141 obtained with the FIES spectrograph available in the 2.56-m Nordic Optical Telescope data archive. The resolving power of the spectrograph in the mode used for these observations is R=67,000. I used iSpec \citep{2014A&A...569A.111B} to measure the radial velocity of the primary star using cross-correlation of the spectra against a numerical mask based on the solar spectrum over the wavelength range 400\,--\,650\,nm. These radial velocities are given in Table~\ref{tab:koi7141_rv}. The value of $K_1$ reported in Table~\ref{tab:basic_info} comes from an unweighted least-squares fit of a Keplerian orbit to these radial velocity measurements with the values of $e=0.7821$, $\omega=328.61$ fixed at the values derived from the fit to the light curve and $T_0$ and $P$ fixed at the values in Table~\ref{tab:ephem}. The fit is shown in Fig.~\ref{fig:KOI-7141_rvfit}. The {\it rms} of the residuals for this fit is 0.3\,km/s.

\begin{table}
\centering
\caption{Barycentric radial velocity of KOI-7141 measured from spectra obtained with the FIES spectrograph. T$_{\rm exp}$ is the exposure time.}
\label{tab:koi7141_rv}
 \begin{tabular}{rrrr} 
 \hline
\multicolumn{1}{l}{BJD$_{\rm TDB}$} & 
\multicolumn{1}{l}{V$_{\rm r}$ [km/s]}  &
\multicolumn{1}{l}{T$_{\rm exp}$ [s]}  & Phase \\
 \hline 
2458455.3513 &$ -2.05 $ &  1380  & 0.843 \\
2458461.3414 &$ 50.35 $ &  1800  & 0.961 \\
2458458.2982 &$ 10.11 $ &  1380  & 0.901 \\
2457909.5511 &$ 25.20 $ &  1380  & 0.022 \\
2458437.3163 &$ 0.48 $ &  1380  & 0.485 \\
2458426.3350 &$ 6.23 $ &  1380  & 0.267 \\
2458459.2987 &$ 33.22 $ &  1380  & 0.921 \\
2457907.6436 &$ 36.56 $ &  1380  & 0.984 \\

\hline 
\end{tabular}
\end{table}



\section{HD 22064}
\label{sec:hd22064}
 There are six high-resolution ($R\approx 22,500$) near-infrared (1514\,--\,1694\,nm) spectra of this star obtained as part of the APOGEE survey \citep{2020AJ....160..120J} available from the Sloan Digital Sky Survey data archive.\footnote{\url{https://www.sdss.org/}} These spectra were obtained on various dates distributed evenly around the spectroscopic orbit. The signal from the M-dwarf companion can be clearly detected in these spectra using the technique described below. This has enabled me to measure accurate, model-independent masses and radii for the stars in this binary, and to directly measure the effective temperature of the primary star.

\subsection{Mass and radius measurements}
 
\begin{figure}
	\includegraphics[width=\columnwidth]{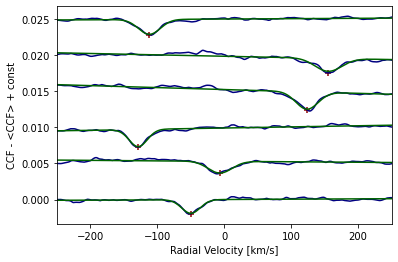}
    \caption{Residual cross-correlation functions computed from APOGEE spectra of HD~22064 after subtracting the signal from the primary star. The red cross shows the radial velocity of the M-dwarf star measured with the Gaussian profile fit by least-squares shown by the green line.}
    \label{fig:HD_22064_ccfs}
\end{figure}

\begin{figure}
	\includegraphics[width=\columnwidth]{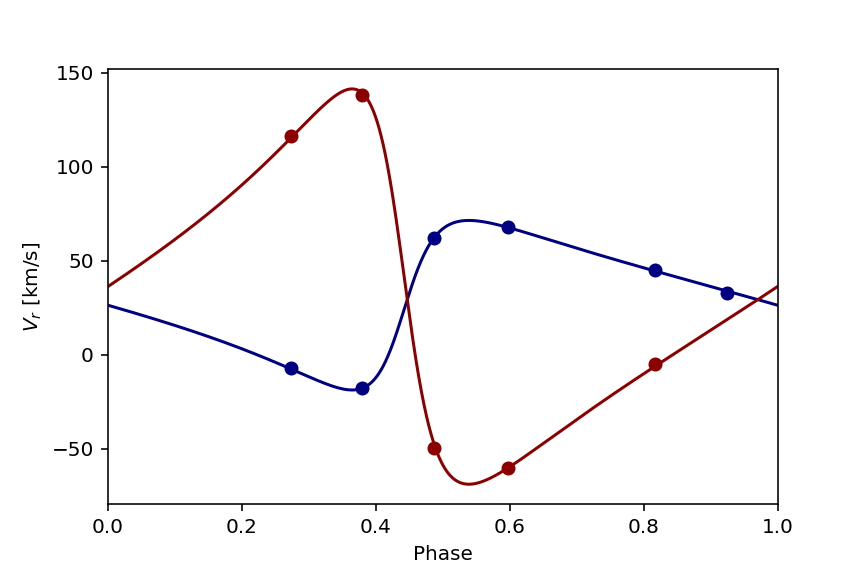}
    \caption{Spectroscopic orbit fits to the radial velocities of HD~22064 measured from APOGEE spectra.}
    \label{fig:HD_22064_rvfit}
\end{figure}

\begin{figure}
	\includegraphics[width=\columnwidth]{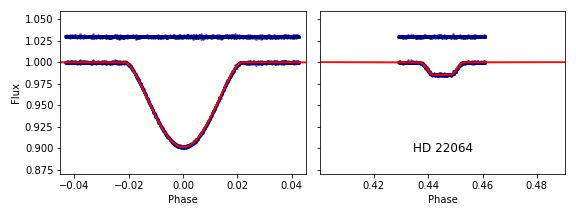}
    \caption{TESS light curve of HD~22064 and residuals from the best fit shown offset vertically in each panel. The width of the two panels in phase units are equal so that the relative widths of the two eclipses can be seen directly. }
    \label{fig:HD_22064_lcfit}
\end{figure}

I used iSpec \citep{2014A&A...569A.111B} to compute the cross-correlation function (CCF) of each APOGEE spectrum against a template spectrum. I used the library of high-resolution synthetic spectra from \citet{2013A&A...553A...6H} to compute (by linear interpolation) a template assuming T$_{\rm eff}=4000$\,K, $\log g= 4.65$ and solar composition. The CCFs are dominated by the signal from the primary star. To remove the signal from the primary star, I shifted the CCFs onto a common velocity scale with the dip from the primary star at $V_r = 0$, computed the mean of these shifted CCFs, and then subtracted this mean CCF from each of the original CCFs. The resulting ``residual CCFs'' are shown in Fig.~\ref{fig:HD_22064_ccfs}, together with the Gaussian profile fit by least-squares that I used to measure the radial velocity of the M-dwarf star. This technique gives reliable radial velocity measurements in the 5 cases where the peaks of the two stars in the CCFs are well separated. The radial velocity of the M-dwarf measured for the one spectrum obtained near conjunction is clearly wrong, so I have removed it from my analysis. For the radial velocity of the primary star, I used the heliocentric radial velocities given in the metadata provided with each spectrum. These radial velocities and the radial velocity of the M-dwarf companion measured relative to them from the residual CCFs are given in Table~\ref{tab:hd22064_rv}.  The spectroscopic orbits measured from these radial velocities are given in Table~\ref{tab:rvfit} and are shown in Fig.~\ref{fig:HD_22064_rvfit}. The priors on $e\cos(\omega)$ and $e\sin(\omega)$ are derived from the least-squares fit to the light curve described in section~\ref{sec:hd22064_lcfit} and shown in Fig.~\ref{fig:HD_22064_lcfit}. The masses and radii of both stars derived from the parameters obtained from the analysis of the light curve and the spectroscopic orbit are given in Table~\ref{tab:hd22064_mrt}.

\begin{table}
\centering
\caption{Heliocentric radial velocities of the stars in the binary system HD~22064 measured from APOGEE spectra. }
\label{tab:hd22064_rv}
 \begin{tabular}{rrrr} 
 \hline
\multicolumn{1}{l}{BJD$_{\rm TDB}$} & 
\multicolumn{1}{l}{V$_{\rm r,1}$ [km/s]}  &
\multicolumn{1}{l}{V$_{\rm r,2}$ [km/s]}  &
\multicolumn{1}{l}{Phase} \\
 \hline 
2457662.96211 &$     44.99  $&$     -4.57 $&  0.816 \\ 
2457663.94745 &$     33.18  $&         --- &  0.924 \\ 
2457660.95990 &$     68.15  $&$    -60.06 $&  0.597 \\ 
2457310.88004 &$     -7.19  $&$    116.49 $&  0.274 \\ 
2457658.96888 &$    -17.73  $&$    137.91 $&  0.379 \\ 
2457659.95935 &$     62.19  $&$    -49.36 $&  0.487 \\ 
\hline 
\end{tabular}
\end{table}

\begin{table}
\centering
\caption{Spectroscopic orbit fit to radial velocities of HD 22064 measured from APOGEE spectra. $T_0$ is the time of primary eclipse and $\sigma_{1,2}$ are the standard deviation of the residuals for each star.}
\label{tab:rvfit}
 \begin{tabular}{lrcrl} 
 \hline
 Parameter & &Value  \\
 \hline 
\noalign{\smallskip}

 \multicolumn{5}{l}{Fixed parameters} \\ 
 $T_0$           &  \multicolumn{3}{r}{$2458578.1295$}& BJD$_{\rm TDB}$  \\
 $P$             & \multicolumn{3}{r}{$ 9.134881$ } & d \\
\noalign{\smallskip}
\multicolumn{5}{l}{Priors} \\
\noalign{\smallskip}
 $e\sin\omega$   & $  -0.5320 $ &$ \pm $ & $  0.0007  $ \\
 $e\cos\omega$   & $-0.0732 $ &$ \pm $ & $ 0.0001  $ \\

 \multicolumn{5}{l}{Free parameters} \\
 $\gamma_1$      & $   14.87 $ &$ \pm $ & $     0.12 $  & km/s \\
 $K_1$           & $   44.98 $ &$ \pm $ & $     0.34 $  & km/s \\
 $\gamma_2$      & $   14.24 $ &$ \pm $ & $     0.45 $  & km/s \\
 $K_2$           & $   105.0 $ &$ \pm $ & $      1.1 $  & km/s \\
 $e\sin\omega$   & $  -0.5320 $ &$ \pm $ & $   0.0007  $ \\
 $e\cos\omega$   & $  -0.0732 $ &$ \pm $ & $   0.0001  $ \\
\noalign{\smallskip}
\multicolumn{5}{l}{Fit statistics} \\
 $\sigma_1$      & &&0.43& km/s \\
 $n_1$      & && 6 &   \\
 $\sigma_2$      & &&1.32& km/s \\
 $n_2$      & && 5 &  \\
 \noalign{\smallskip}
 \multicolumn{5}{l}{Derived parameters} \\
 $e$             & $  0.5370 $ &$ \pm $ & $   0.0007  $ \\
 $\omega$        & $262\fdg17 $ &$ \pm $ & $    0\fdg02  $ \\\hline
\end{tabular}
\end{table}

\begin{table}
\caption[]{Mass, radius, effective temperature and derived parameters for the stars in HD~22064. N.B. $T_{\rm eff,1}$  and $T_{\rm eff,2}$ are subject to additional systematic uncertainty of 10\,K and 8\,K, respectively.}
\label{tab:hd22064_mrt}
\begin{center}
  \begin{tabular}{lrrr}
\hline
\noalign{\smallskip}
 \multicolumn{1}{l}{Parameter} &
 \multicolumn{1}{l}{Value} &
 \multicolumn{1}{l}{Error} &
 \multicolumn{1}{r}{} \\
\noalign{\smallskip}
\hline
\noalign{\smallskip}
$M_1/{\mathcal M^{\rm N}_{\odot}}$&1.345  & $\pm$ 0.031 & [2.3\%] \\
\noalign{\smallskip}
$M_2/{\mathcal M^{\rm N}_{\odot}}$&0.576 & $\pm$ 0.010 &[1.8\%] \\
\noalign{\smallskip}
$R_1/{\mathcal R^{\rm N}_{\odot}}$&1.554 & $\pm$ 0.014 &[0.9\%] \\
\noalign{\smallskip}
$R_2/{\mathcal R^{\rm N}_{\odot}}$&0.595 & $\pm$ 0.008 &[1.4\%] \\
\noalign{\smallskip}
$T_{\rm eff,1} $ [K]  & $ 6763 $ & $\pm$ 39  & [0.6\%] \\
\noalign{\smallskip}
$T_{\rm eff,2} $ [K]  & $ 3700 $ & $\pm$ 315  & [8.5\%] \\
\noalign{\smallskip}
$\rho_1/{\rho^{\rm N}_{\odot}}$  &0.359 & $\pm$ 0.006 &[1.6\%] \\
\noalign{\smallskip}
$\rho_2/{\rho^{\rm N}_{\odot}}$  & 2.74 & $\pm$ 0.10 &[3.6\%] \\
\noalign{\smallskip}
$\log g_1$ [cgs] & 4.184 & $\pm$ 0.006 &   \\
\noalign{\smallskip}
$\log g_2$ [cgs] & 4.65 & $\pm$ 0.01 &  \\
\noalign{\smallskip}
$\log L_1/{\mathcal L^{\rm N}_{\odot}}$   & 0.658 & $\pm$ 0.006 &   \\
\noalign{\smallskip}
$\log L_2/{\mathcal L^{\rm N}_{\odot}}$   & $ -1.176 $& $\pm$ 0.012 &   \\
\noalign{\smallskip}
[M/H]   & $-0.05$ & $\pm 0.15$ & \\
\noalign{\smallskip}
[Si/Fe] & $+0.3$  & $\pm 0.15$ & \\
\noalign{\smallskip}
[Mg/Fe] & $+0.3$  & $\pm 0.15$ & \\
\noalign{\smallskip}
\hline
\end{tabular}
\end{center}
\end{table}

\subsection{Direct measurement of the stellar effective temperature}
The effective temperature for a star with Rosseland radius $R$ and total luminosity $L$ is defined by the equation  
$$L=4\pi R^2 \sigma_{\rm SB} {\rm T}_{\rm eff}^4,$$
where $\sigma_{\rm SB}$ is the Stefan-Boltzmann constant. For a binary star at distance $d$, i.e. with parallax $\varpi=1/d$, the flux corrected for extinction observed at the top of Earth's atmosphere is
$$f_{0,b}= f_{0,1}+f_{0,2}=\frac{\sigma_{\rm SB}}{4}\left[\theta_1^2{\rm T}_{\rm eff,1}^4 + \theta_2^2{\rm T}_{\rm eff,2}^4\right],$$
where $\theta_1=2R_1\varpi$ is the angular diameter of star 1, and similarly for star 2. All the quantities are known or can be measured for HD~22064 provided we can accurately integrate the observed flux distributions for the two stars independently. This is possible because photometry of the combined flux from both stars is available from ultraviolet to mid-infrared wavelengths, and the flux ratio has been measured from the TESS light curve. Although we have no direct measurement of the flux ratio at infrared wavelengths, we can make a reasonable estimate for the small contribution of the M-dwarf to the measured total infrared flux using empirical colour\,--\,$T_{\rm eff}$ relations. The M-dwarf contributes less than 1.5\,per~cent to the total flux so it is not necessary to make a very accurate estimate of the M-dwarf flux distribution in order to derive an accurate value of $T_{\rm eff}$ for the primary star.

The photometry used in this analysis is given in Table~\ref{tab:mags}. The NUV and FUV magnitudes are taken from GALEX data release GR7 \citep{2014yCat.2335....0B} with a correction to the IUE flux scale based on the results from \citet{2014MNRAS.438.3111C}. We assume that the flux from the M-dwarf at ultraviolet wavelengths is negligible. The Gaia photometry is from Gaia data release EDR3. J, H and Ks magnitudes are from the 2MASS survey \citep{2006AJ....131.1163S}. Unfortunately, the 2MASS observations were obtained during a primary eclipse. I subtracted 0.1 magnitudes from the catalogued 2MASS J, H and Ks magnitudes to account for this and assumed a standard error of 0.05 magnitudes for these corrected magnitudes. WISE magnitudes are from the All-Sky Release Catalog \citep{2012yCat.2311....0C} with corrections to Vega magnitudes made as recommended by \citet{2011ApJ...735..112J}. Photometry in the PanSTARRS-1 photometry system is taken from \citet{2018ApJ...867..105T}. Details of the zero-points and response functions used to calculate synthetic photometry from an assumed spectral energy distribution are given in \citet{2020MNRAS.497.2899M}. Tycho B$_{\rm T}$ and V$_{\rm T}$ magnitudes are taken from \citet{2000A&A...355L..27H}. Zero-points and photonic passbands for these magnitudes are taken from \citet{2012PASP..124..140B}. Analysis of Str\"{o}mgren photometry for HD~22064 by \citet{1996PASP..108..772P} shows that the reddening towards this star is very low, so I have used a Gaussian prior ${\rm E}({\rm B}-{\rm V}) = 0.00 \pm 0.01$ in my analysis.  

To establish colour\,--\,$T_{\rm eff}$ relations suitable for dwarf stars with $3400\,{\rm K} < {\rm T}_{\rm eff} < 4200\,{\rm K} $ we use a robust linear fit to the stars listed in Table~5  of \citet{2015ApJ...804...64M} within this $T_{\rm eff}$ range. Photometry for these stars is taken from the TESS input catalogue. To estimate a suitable standard error for a Gaussian prior based on this fit we use 1.25$\times$ the mean absolute deviation of the residuals from the fit. Colour\,--\,$T_{\rm eff}$ relations suitable for the primary star were calculated in similar way based on stars selected from the Geneva-Copenhagen survey \citep{2009A&A...501..941H, 2011A&A...530A.138C}  with $6500\,{\rm K} < {\rm T}_{\rm eff} < 7000\,{\rm K}$, $E({\rm B}-{\rm V})<0.01$ and $3.5 < \log g < 4.5$. The results are given in Table~\ref{tab:frp}.

\begin{table*}
\centering
\caption{Observed apparent magnitudes for HD~22064 and predicted values based on our synthetic photometry. The predicted magnitudes are shown with error estimates from the uncertainty on the zero-points for each photometric system. The pivot wavelength for each band pass is shown in the column headed $\lambda_{\rm pivot}$. The magnitudes of the primary star alone corrected for the contribution to the total flux from the M-dwarf are shown in the column headed $m_1$. The flux ratio in each band is shown in the final column.}
\label{tab:mags}
\begin{tabular}{lrrrrrr} 
\hline
		Band &  $\lambda_{\rm pivot}$ [nm]& \multicolumn{1}{c}{Observed} & \multicolumn{1}{c}{Computed} & 
\multicolumn{1}{c}{$\rm O-\rm C$} & \multicolumn{1}{c}{$m_1$} & \multicolumn{1}{c}{$\ell$} [\%]\\
\hline
FUV &   1\,535 &$ 17.237 \pm 0.391 $&$ 17.721 \pm 0.134 $&$ -0.484 \pm 0.413 $&$  17.237 \pm 0.391 $&  0.00 \\
NUV &   2\,301 &$ 12.733 \pm 1.222 $&$ 12.523 \pm 0.154 $&$ +0.210 \pm 1.232 $&$  12.733 \pm 1.222 $&  0.00 \\
u   &   3\,493 &$ 10.125 \pm 0.006 $&$ 10.176 \pm 0.200 $&$ -0.051 \pm 0.200 $&$  10.125 \pm 0.006 $&  0.05 \\
v   &   3\,836 &$  9.674 \pm 0.008 $&$  9.670 \pm 0.100 $&$ +0.004 \pm 0.100 $&$   9.675 \pm 0.008 $&  0.08 \\
BT  &   4\,190 &$  9.291 \pm 0.020 $&$  9.332 \pm 0.040 $&$ -0.041 \pm 0.045 $&$   9.293 \pm 0.020 $&  0.16 \\
BP  &   5\,110 &$  8.951 \pm 0.003 $&$  8.952 \pm 0.003 $&$ -0.001 \pm 0.004 $&$   8.956 \pm 0.003 $&  0.46 \\
VT  &   5\,300 &$  8.860 \pm 0.016 $&$  8.889 \pm 0.040 $&$ -0.029 \pm 0.043 $&$   8.865 \pm 0.016 $&  0.44 \\
G   &   6\,218 &$  8.752 \pm 0.003 $&$  8.747 \pm 0.003 $&$ +0.005 \pm 0.004 $&$   8.762 \pm 0.003 $&  0.90 \\
RP  &   7\,769 &$  8.401 \pm 0.004 $&$  8.403 \pm 0.004 $&$ -0.002 \pm 0.005 $&$   8.418 \pm 0.004 $&  1.55 \\
J   &  12\,406 &$  8.072 \pm 0.104 $&$  8.023 \pm 0.005 $&$ +0.049 \pm 0.104 $&$   8.111 \pm 0.104 $&  3.54 \\
H   &  16\,494 &$  7.906 \pm 0.104 $&$  7.861 \pm 0.005 $&$ +0.045 \pm 0.104 $&$   7.965 \pm 0.104 $&  5.26 \\
Ks  &  21\,638 &$  7.825 \pm 0.104 $&$  7.809 \pm 0.005 $&$ +0.016 \pm 0.104 $&$   7.889 \pm 0.104 $&  5.69 \\
W1  &  33\,682 &$  7.750 \pm 0.024 $&$  7.787 \pm 0.002 $&$ -0.037 \pm 0.024 $&$   7.813 \pm 0.024 $&  5.68 \\
W2  &  46\,179 &$  7.789 \pm 0.019 $&$  7.788 \pm 0.002 $&$ +0.001 \pm 0.019 $&$   7.854 \pm 0.019 $&  5.85 \\
W3  & 120\,731 &$  7.804 \pm 0.019 $&$  7.790 \pm 0.002 $&$ +0.014 \pm 0.019 $&$   7.868 \pm 0.020 $&  5.74 \\
W4  & 221\,942 &$  8.205 \pm 0.232 $&$  7.826 \pm 0.002 $&$ +0.379 \pm 0.232 $&$   8.286 \pm 0.232 $&  7.22 \\
\hline
\end{tabular}
\end{table*}

\begin{table}
	\centering
	\caption{Colour-$T_{\rm eff}$ relations used to establish Gaussian priors on the flux ratio at infrared wavelengths for HD~22064. The dependent variables are $X_1 = T_{\rm eff,1}-6.75\,{\rm kK}$ and $X_2 = T_{\rm eff,2}-4.0\,{\rm kK}$. }
	\label{tab:frp}
	\begin{tabular}{lll} 
		\hline
		Colour &  Primary & Secondary  \\
		\hline
V$-$J  & $ 0.777 -0.3865 \, X_1 \pm  0.016 $ & $  2.799 -0.0015 \, X_2 \pm  0.107 $ \\
V$-$H  & $ 0.928 -0.5152 \, X_1 \pm  0.018 $ & $  3.466 -0.0014 \, X_2 \pm  0.111 $ \\
V$-$Ks & $ 0.989 -0.5276 \, X_1 \pm  0.016 $ & $  3.662 -0.0015 \, X_2 \pm  0.131 $ \\
V$-$W1 & $ 1.031 -0.5437 \, X_1 \pm  0.025 $ & $  3.686 -0.0015 \, X_2 \pm  0.105 $ \\
V$-$W2 & $ 1.046 -0.5173 \, X_1 \pm  0.039 $ & $  3.732 -0.0018 \, X_2 \pm  0.120 $ \\
V$-$W3 & $ 0.990 -0.5221 \, X_1 \pm  0.021 $ & $  3.699 -0.0018 \, X_2 \pm  0.121 $ \\
V$-$W4 & $ 1.046 -0.4618 \, X_1 \pm  0.050 $ & $  3.795 -0.0017 \, X_2 \pm  0.131 $ \\
\hline
\end{tabular}
\end{table}

The method we have developed to measure $T_{\rm eff}$ for eclipsing binary stars is described fully in \citet{2020MNRAS.497.2899M}. Briefly, we use  {\sc emcee} \citep{2013PASP..125..306F} to sample the posterior probability distribution $P(\Theta| D)\propto P(D|\Theta)P(\Theta)$ for the model parameters $\Theta$ with prior $P(\Theta)$ given the data, $D$ (observed apparent magnitudes and flux ratios). The model parameters are  $$\Theta = \left({\rm T}_{\rm eff,1},  {\rm T}_{\rm eff,2}, \theta_1, \theta_2, {\rm E}({\rm B}-{\rm V}), \sigma_{\rm ext}, \sigma_{\ell},  c_{1,1}, \dots, c_{2,1}, \dots\right).$$  The prior $P(\Theta)$ is calculated using the angular diameters $\theta_1$ and $\theta_2$ derived from the radii $R_1$ and $R_2$ and the parallax $\varpi$, the priors on the flux ratio at infrared wavelengths based on the colour\,--\,T$_{\rm eff}$ relations in Table~\ref{tab:frp}, and the Gaussian prior on the reddening described above. The hyperparameters $\sigma_{\rm ext}$ and $\sigma_{\ell}$ account for additional uncertainties in the synthetic magnitudes  and flux ratio, respectively, due to errors in zero-points, inaccurate response functions, stellar variability, etc. The parallax is taken from Gaia EDR3 with corrections to the zero-point from \citet{2022MNRAS.509.4276F}. 

To calculate the synthetic photometry for a given value of $T_{\rm eff}$ we used a model spectral energy distribution (SED) multiplied by a distortion function, $\Delta(\lambda)$. The distortion function is a linear superposition of Legendre polynomials in log wavelength. The coefficients of the distortion function for star 1 are $c_{1,1}, c_{1,2}, \dots$, and similarly for star 2. The distorted SED for each star is normalized so that the total apparent flux prior to applying reddening is $\sigma_{\rm SB}\theta^2{\rm T}_{\rm eff}^4/4$. These distorted SEDs provide a convenient function that we can integrate to calculate synthetic photometry that has realistic stellar absorption features, and where the overall shape can be adjusted to match the observed magnitudes from ultraviolet to infrared wavelengths, i.e. the effective temperatures we derive are based on the integrated stellar flux and the star's angular diameter, not SED fitting.

For this analysis we use model SEDs computed from BT-Settl model atmospheres \citep{2013MSAIS..24..128A} obtained from the Spanish Virtual Observatory.\footnote{\url{http://svo2.cab.inta-csic.es/theory/newov2/index.php?models=bt-settl}} We use linear interpolation to obtain a reference SED for the primary star appropriate for  $T_{\rm eff,1}=6750$\,K, $\log g_1 = 4.18$, $[{\rm Fe/H}] = 0.0$ and   $[{\rm \alpha/Fe}] = 0.0$. For the reference SED for the M dwarf companion we assume $T_{\rm eff,1}=4000$\,K, $\log g_1 = 4.65$, and the same composition. A similar analysis assuming the value of [Fe/H] from Table~\ref{tab:basic_info} gave a poor fit to the observations and an estimate of T$_{\rm eff,2}$ that is far too low for a star with a mass $\approx 0.5 M_{\odot}$.  
We experimented with distortion functions with 1, 2 or 3 coefficients per star and found the results to be very similar in all cases. The results presented here use one distortion coefficient per star because there is no improvement in the quality if the fit if we use a larger number of coefficients. The predicted apparent  magnitudes including their uncertainties from errors in the zero-points for each photometric system are compared to the observed apparent magnitudes in Table~\ref{tab:mags}. The effective temperatures derived are given in Table~\ref{tab:hd22064_mrt}. The random errors quoted in Table~\ref{tab:hd22064_mrt} do not allow for the systematic error due to the uncertainty in the absolute calibration of the CALSPEC flux scale \citep{2014PASP..126..711B}. This additional systematic error is 10\,K for the  primary star and 8\,K for the M-dwarf companion. 

\begin{figure}
	\includegraphics[width=\columnwidth]{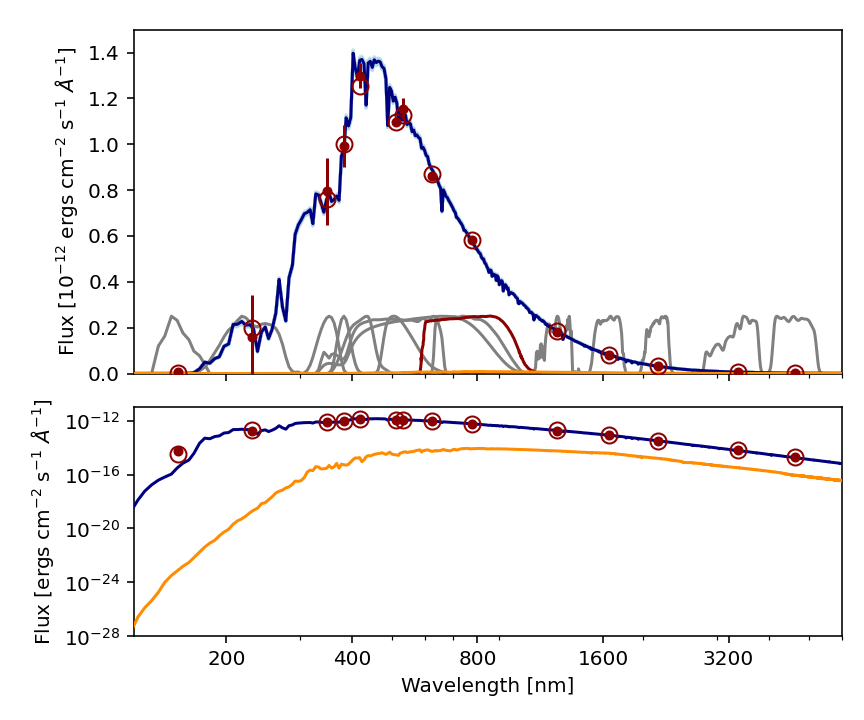}
    \caption{The spectral energy distribution (SED) of HD~22064. The observed fluxes are plotted with open circles and the predicted fluxes for the mean of the posterior probability distribution (PPD) integrated over the response functions shown in grey are plotted with filled symbols. The SED predicted by the mean of the PPD is plotted in dark blue and light blue shows the SEDs produced from 100 random samples from the PPD. The contribution to the total SED from the M dwarf is shown in orange. The W3 and W4 mid-infrared bands also used in the analysis are not shown here. The TESS photometric band is shown in dark red. }
    \label{fig:sed}
\end{figure}

\subsection{Analysis of the APOGEE spectrum}
 I used synthetic SEDs from \citet{2013A&A...553A...6H} to estimate the mean flux ratio in the APOGEE spectral range assuming that the flux ratio in the TESS bandpass is $\ell_{\rm T} = 0.0147$, T$_{\rm eff, 1} = 6800$\,K, $\log g_1=4.2$, T$_{\rm eff, 2} = 3800$\,K, $\log g_2=4.65$, and solar composition. The mean flux ratio in the APOGEE band is then estimated to be $\ell_{\rm APOGEE} \approx 0.044$ with an appreciable slope of about 0.00005\,nm$^{-1}$. \citet{2020AJ....160..120J} report ${\rm [Fe/H]}=-0.12$ from their analysis of five co-added APOGEE spectra.  They did not account for the dilution of the spectral lines from the primary star by the M-dwarf. This is likely to lead to a slight underestimate of the surface chemical abundances estimated from the analysis of this spectrum, which suggests that the metallicity of HD~22064 is approximately solar.

 I have reanalysed the APOGEE spectra of HD~22064 after removing the contribution from the M-dwarf companion. To estimate the contribution of the M-dwarf to APOGEE spectrum, I subtracted the template used to compute the CCFs for a range of trial flux ratios and then measured the depth of the peak in the residual CCFs by fitting a Gaussian profile at the radial velocity expected based on the spectroscopic orbit fit, and with the width fixed at the mean value measured from the fit to the uncorrected residual CCFs. The slope in the flux ratio estimated above was accounted for before the scaled template spectrum was subtracted from the observed spectra. I also found that broadening the spectral lines in the template assuming a projected rotational velocity $v_{\rm rot}\sin i \approx 15$\,km\,s$^{-1}$ was needed to accurately remove the signal of the M-dwarf from the residual CCFs. From these results, I estimate that the actual flux ratio in the APOGEE bandpass is $\ell_{\rm APOGEE} = 0.0404 \pm 0.0018$, which agrees well with my estimate based on the synthetic SEDs from \citet{2013A&A...553A...6H}. The APOGEE spectrum of the primary star after removal of the M-dwarf spectrum from the combined spectra assuming this flux ratio is shown in Fig.~\ref{fig:apogee_spectrum}. These data are provided in the supplementary online information that accompanies this article.  

 For the analysis of this primary star spectrum, I used the {\sc spectrum} spectral synthesis code \citep{1994AJ....107..742G} with the atomic line data for the APOGEE spectral range and the MARCS model stellar atmosphere structure data provided with iSpec \citep{2014A&A...569A.111B}. The assumed solar chemical composition is taken from \citet{2007SSRv..130..105G}. Broad spectral features such as absorption lines due to hydrogen are badly affected by the processing steps leading to the production of the observed spectrum shown in Fig.~\ref{fig:apogee_spectrum}. This makes it difficult to directly compare synthesised stellar spectra to the observed spectrum. Instead, I compute the ratio of the observed spectrum to the synthesised spectrum, and then used a smoothed version of this ratio as a model for the systematic errors in the broad spectral features in the observed spectrum. This enabled me to compute a ``reconstructed spectrum'' by multiplying the observed primary star spectrum by the smoothed version of the ratio. The smoothing is done using a Savitzky–Golay filter with a width of 2.9\,nm so that the equivalent widths of narrow absorption lines are negligibly affected. Note that the cores of the hydrogen Brackett absorption lines are not accurately reconstructed by this method. The assumed effective temperature and surface gravity were fixed at the values T$_{\rm eff} = 6750$\,K and $\log g = 4.18$. Rotational broadening $v_{\rm rot}\sin i =25.5$\,km\,s$^{\-1}$  was estimated by measuring the width of the cross-correlation function of the observed spectrum against the spectrum of the Sun, and calibrating the relation between $v_{\rm rot}\sin i$ and CCF width using synthetic spectra. I then computed $\chi^2$ for the model spectrum compared to the reconstructed spectrum assuming a range of [Fe/H] values and scaled solar abundances. I found that the quality of the fit judged by-eye was better if I increased the abundances of Mg and Si by +0.3\,dex relative to scaled solar abundances. Repeating the analysis with these enhanced Mg and Si abundances, I found that the best-fit metallicity occurs for ${\rm [M/H]} = -0.03$. Based on the results from independent analyses of APOGEE spectra by \citet{2018AJ....156..126J}, we assume an accuracy of 0.15\,dex for these abundance estimates. The best-fit model spectrum and reconstructed primary star spectrum are shown in Fig.~\ref{fig:apogee_spectrum}. These data are also provided in the supplementary online information that accompanies this article. 
 
 \subsection{Discussion}
  
\begin{figure}
	\includegraphics[width=0.9\columnwidth]{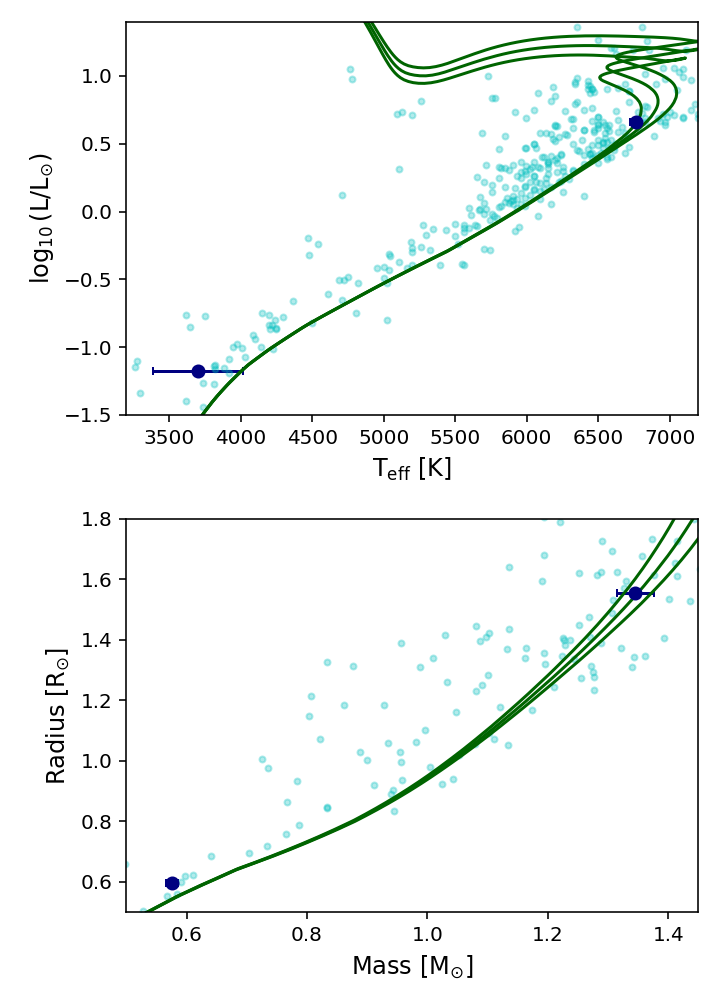}
    \caption{HD 22064 in the mass-radius plane and Hertzsprung-Russell diagrams compared to isochrones for ages of  $1.8\pm0.3$\,Gyr assuming $[{\rm Fe/H}]=0.0$ from  MIST (green dashed line). Pale blue points are data for stars in detached eclipsing binaries from DEBCat \citep{2015ASPC..496..164S}.}
    \label{fig:HRD}
\end{figure}

 The location of HD~22064 in the mass-radius plane and Hertzsprung-Russell diagrams compared to isochrones from MESA Isochrones \& Stellar Tracks \citep[MIST, ][]{2016ApJ...823..102C} is shown in Fig.~\ref{fig:HRD}. The properties of the primary star are well matched by isochrones for an age of about 2\,Gyr. The M-dwarf companion is larger and cooler than predicted by stellar models for star of its mass at this age. This is also seen for other low-mass stars in eclipsing binaries. This persistent discrepancy between models and observations for the properties of low-mass stars is known as the radius inflation problem \citep{2023MNRAS.519.3546S, 2008MNRAS.389..585C, 2013ApJ...776...87S, 2018AJ....155..225K}.
 
\begin{figure}
	\includegraphics[width=\columnwidth]{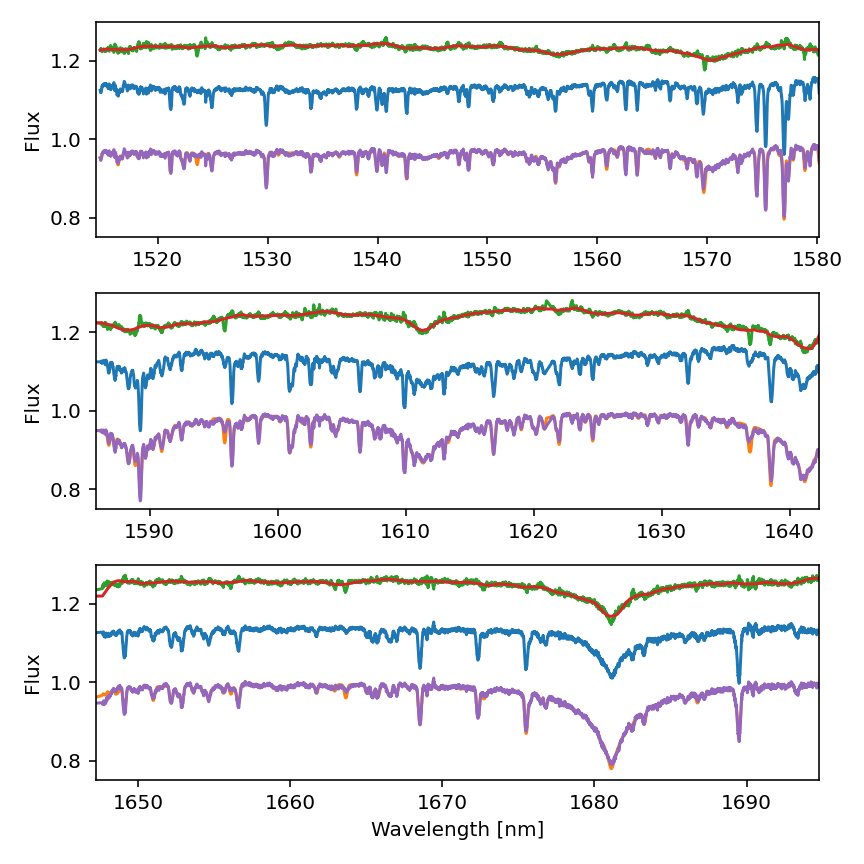}
    \caption{The APOGEE spectrum of the primary star in HD~22064 after removal of the M-dwarf spectrum from the combined spectra. From top to bottom: ratio of the observed spectrum to the best-fit model spectrum and a smoothed version of this ratio, both offset by +0.25 units; observed spectrum of the primary star corrected for the flux from the M-dwarf companion offset by 0.15 units; reconstructed spectrum of the primary star and the best-fit model spectrum.}
    \label{fig:apogee_spectrum}
\end{figure}

\section{Conclusions}

I have identified and characterised a sample of 20 eclipsing binary stars with flux ratios at optical wavelengths $\approx 1$\,per~cent. These will be excellent benchmark stars for large-scale spectroscopic surveys and for the PLATO mission once they have been fully characterised using high-resolution spectroscopy at near-infrared wavelengths to measure the radial velocity of their M-dwarf companions. I have used archival near-infrared spectroscopy for HD~22064 to demonstrate the feasibility of such follow-up observations, and to show that a precision of 50\,K or better is possible for direct T$_{\rm eff}$ measurements of such eclipsing binaries using archival photometry and Gaia parallax measurements. The analysis methods used for this binary also demonstrate the feasibility of effectively removing the signal of the M-dwarf from the combined spectrum of the binary, leaving a clean spectrum of the primary star suitable for detailed spectroscopic analysis. This sample of stars will also be useful for better understanding the radius inflation problem for low-mass stars. 

\section*{Acknowledgements}

PM thanks the referee for their comments on the manuscript that have improved the quality and accuracy of the results presented herein. PM also thanks John Southworth for providing comments and corrections on the manuscript. 

This research was supported by UK Science and Technology Facilities Council (STFC) research grant number ST/M001040/1.

This research made use of Lightkurve, a Python package for Kepler and TESS data analysis \citep{2018ascl.soft12013L}.

This paper includes data collected by the Kepler and TESS missions obtained from the MAST data archive at the Space Telescope Science Institute (STScI). Funding for the Kepler mission is provided by the NASA Science Mission Directorate. Funding for the TESS mission is provided by the NASA Explorer Program. STScI is operated by the Association of Universities for Research in Astronomy, Inc., under NASA contract NAS 5–26555.

This research has made use of the VizieR catalogue access tool, CDS, Strasbourg, France (DOI : 10.26093/cds/vizier). 

 Funding for the Sloan Digital Sky 
Survey IV has been provided by the 
Alfred P. Sloan Foundation, the U.S. 
Department of Energy Office of 
Science, and the Participating 
Institutions. 

SDSS-IV acknowledges support and 
resources from the Center for High 
Performance Computing  at the 
University of Utah. The SDSS 
website is www.sdss.org.

SDSS-IV is managed by the 
Astrophysical Research Consortium 
for the Participating Institutions 
of the SDSS Collaboration including 
the Brazilian Participation Group, 
the Carnegie Institution for Science, 
Carnegie Mellon University, Center for 
Astrophysics | Harvard \& 
Smithsonian, the Chilean Participation 
Group, the French Participation Group, 
Instituto de Astrof\'isica de 
Canarias, The Johns Hopkins 
University, Kavli Institute for the 
Physics and Mathematics of the 
Universe (IPMU) / University of 
Tokyo, the Korean Participation Group, 
Lawrence Berkeley National Laboratory, 
Leibniz Institut f\"ur Astrophysik 
Potsdam (AIP),  Max-Planck-Institut 
f\"ur Astronomie (MPIA Heidelberg), 
Max-Planck-Institut f\"ur 
Astrophysik (MPA Garching), 
Max-Planck-Institut f\"ur 
Extraterrestrische Physik (MPE), 
National Astronomical Observatories of 
China, New Mexico State University, 
New York University, University of 
Notre Dame, Observat\'ario 
Nacional / MCTI, The Ohio State 
University, Pennsylvania State 
University, Shanghai 
Astronomical Observatory, United 
Kingdom Participation Group, 
Universidad Nacional Aut\'onoma 
de M\'exico, University of Arizona, 
University of Colorado Boulder, 
University of Oxford, University of 
Portsmouth, University of Utah, 
University of Virginia, University 
of Washington, University of 
Wisconsin, Vanderbilt University, 
and Yale University.

Based on observations made with the Nordic Optical Telescope, owned in collaboration by the University of Turku and Aarhus University, and operated jointly by Aarhus University, the University of Turku and the University of Oslo, representing Denmark, Finland and Norway, the University of Iceland and Stockholm University at the Observatorio del Roque de los Muchachos, La Palma, Spain, of the Instituto de Astrofisica de Canarias.

\section*{Data Availability}

The data underlying this article are available from the following sources: MAST data archive at the Space Telescope Science Institute (STScI) (\url{https://archive.stsci.edu}); VizieR catalogue access tool hosted by the Centre de Données astronomiques de Strasbourg (\url{https://vizier.cds.unistra.fr/}); Sloan Digital Sky Survey (SDSS) Science Archive Server (SAS, \url{https://www.sdss.org/}).



\bibliographystyle{mnras}
\bibliography{all} 







\bsp	
\label{lastpage}
\end{document}